# Information Coefficient as a Performance Measure of Stock Selection Models

Feng Zhang, Ruite Guo and Honggao Cao[1]
Wells Fargo & Company
October, 2020

**Abstract**

Information coefficient (IC) is a widely used metric for measuring investment managers' skills in selecting stocks. However, its adequacy and effectiveness for evaluating stock selection models has not been clearly understood, as IC from a realistic stock selection model can hardly be materially different from zero and is often accompanied with high volatility. In this paper, we investigate the behavior of IC as a performance measure of stock selection models. Through simulation and simple statistical modeling, we examine the IC behavior both statically and dynamically. This examination helps us propose two practical procedures that one may use for IC-based ongoing performance monitoring of stock selection models.

## 1. Introduction

Information coefficient (IC) is a widely used metric for measuring investment managers' skills in selecting stocks. Statistically, this metric describes the correlation between the actual and predicted stock returns that the managers considered in the investment management process. Because the predicted returns typically come from a stock selection model (or "alpha model"), information coefficient is also often used as a measure for the performance of such a model. The higher the information coefficient, the better performing the underlying stock selection model, and the more skillful the managers or model developers, suggests the conventional wisdom.

But the metric has not been clearly understood or carefully investigated in terms of its adequacy and effectiveness for evaluating real stock selection models. On the one hand, the textbook definition of IC suggests that a perfect stock picker would have an IC = 1, whereas a terrible picker might end up with an IC = -1, implying that the full spectrum of the correlation distribution can be used for model performance evaluation. On the other hand, because of the inherent difficulties of predicting stock returns, it is well accepted that a realistic stock selection model can hardly have an IC materially different from 0 ( [1], [2]). A stellar model, for example, may often have an IC of 0.05 or 0.1, or otherwise "barely" above zero; to become a very lousy model, the IC needs not to be deeply into the negative territory. In short, the real IC distribution tends to be narrow, concentrating around zero.

The disconnect between the textbook view of the IC distribution and its reality in the investment field creates a number of unique problems when using IC to evaluate stock selection models. First, because real ICs are rarely materially different from zero, hypothesis testing on the information contained in a model becomes more difficult. Statistics and materiality may collide in producing inconsistent messages about the performance of a model. A model with an IC = 0.01 can be said to contain weak information (materiality), but the information may not be easily detected (statistics). Conversely, a realized small but positive IC such as 0.01 may not necessarily indicate the model contains real information.

---

[1] Opinions expressed in this paper do not necessarily reflect those of Wells Fargo & CO. We thank Dingxi Qiu, Harindra de Silva and Wai Lee for valuable comments on an earlier version of this paper. All errors are our own, however. Correspondence should be addressed to Honggao Cao at *honggao.cao@wellsfargo.com*.

Second, small real ICs requires a different paradigm to look at the correlations, *per se*. As "weak" correlations are prevalent in stock selection models, new criteria need to be established to enable a magnifying view of the correlations, helping distinguish between meaningful models and "placebos".

Finally, because stock selection models are always measured against random actual returns, ICs are generally a random variable ( [1]). The performance of a stock selection model – and that of the investments supported by the model – is thus ultimately determined by how this small, random variable behaves across time.

In this paper, we investigate the behavior of IC as a metric measuring the performance of stock selection models. Our study can be divided into two parts. In the first part, we examine the IC behavior statically, trying to understand it from a snapshot perspective. We start by presenting a stylized framework to link IC as a performance measure of a stock selection model. It shows how the actual and predicted stock returns are related to each other in a classical simple regression setting. This framework will help illustrate some unique aspects of the IC problems. We then conduct a set of simulation tests to examine the behavior of the "small" ICs. We analyze the statistical properties of the simulated ICs, and discuss how these properties may be used when we examine realized ICs and determine acceptable IC levels for models.

In the second part, which starts in section 7, we examine the realized ICs dynamically, leveraging data from some industry studies. We also discuss two practical procedures that use the realized ICs for the ongoing model performance monitoring.

It should be clearly noted that IC is just one of the performance measures used for stock selection models. Other popular measures including quintile spread, hit ratio, and batting average are often used in tandem with IC. In addition, the performance of a stock selection model should not be confused with the performance of the investments supported by the model. While managers are increasingly relying on quantitative tools in crafting their investment decisions, the performance of the investments is also influenced by qualitative strategies, processes and discretions, as may be reflected in the "breadth" term of the "fundamental law of active asset management" ( [3]).

## 2. Information Coefficient in Simple Regression

Consider an investment universe with *N* stocks. The "normalized" actual returns at time *t* can be defined as a random vector $y_t$. An "exogenous" stock selection model[2] predicts the normalized returns at *t* for the stocks in the universe as $x_t$. A simple regression of the actual returns on the predicted returns can then be expressed as in equation (1):

$$y_t = \alpha_t x_t + \varepsilon_t \quad (1)$$

Here, $E(y_t) = 0$, $\sigma(y_t) = 1$, $E(x_t) = 0$ and $\sigma(x_t) = 1$ by the "normalization" of the returns, and $\varepsilon_t$ is a vector of random errors, which is presumably orthogonal to $x_t$.

In this setup, the regression coefficient, $\alpha_t$, is the information coefficient (IC) of the stock selection model. To see this, note:

$$\alpha = \frac{COV(x,y)}{V_x} = \frac{COV(x,y)}{\sigma_x \sigma_x} = \frac{COV(x,y)}{\sigma_x \sigma_y} * \frac{\sigma_y}{\sigma_x} = IC * 1 \quad (2)$$

---

[2] Strictly speaking, a stock-selection model is not "exogenous" as it is typically developed and/or recalibrated with the historical information about the stocks in an investment universe. But all such models can be seen as "exogenous" for our purposes here as how the models are developed is irrelevant to our discussion.

All the subscripts are suppressed for the ease of the presentation.

Also note that the regression coefficient (and thus the IC) is indexed with a "$t$". This does not mean that the coefficient is a variable in this regression; we index it here to set the stage for studying the dynamic behavior of the IC later.

A low IC from a stock selection model implies that $\alpha_t$ is small, close to zero; it also means that the R-squared from the regression tends to be close to zero. Traditionally, one would see this as evidence that the stock selection model is very weak or even irrelevant. But the low IC and low R-squared is a fact, indicative of the inherent difficulties in predicting stock returns. A more relevant question about the IC is thus how to decode the realized low ICs and identify from them potential information and/or problems in the stock selection model.

## 3. Simulation Design

We use simulation analysis to study the behavior of ICs. We consider several factors when designing simulations. The first is the size of the investment universe. A typical stock selection model is often tied to a specific investment universe such as a well-established index or benchmark. SP500, Russell3000, and Stoxx Europe 50 and Stoxx Global 1600 are some examples of these benchmarks. A stock selection model aims to rank order the stocks in the universe in terms of their relative returns. When the model-based returns show weak correlations with actual returns, the credibility of the model is affected by the universe size.

The second factor is the distribution of the stock returns in the investment universe. Different from the return distribution of individual stocks (e.g., Googles GOOG, Wells Fargo WFC and Alibaba BABA) across time, this distribution is about the behavior of the stocks in the universe cross-sectionally. We find that a normal distribution can be a good approximation for our purposes in this paper[3], although further research may be needed on the topic.

A third factor is the IC range. Although our focus is on the behavior of low ICs, how low is low? How differently is the IC behaving near zero and away from zero? Because IC is often close to zero, would a negative IC ever be acceptable? Eventually, we settle this problem after going through some preliminary exploration.

A fourth factor is about the definition of the correlation used for IC. Should the correlation be calculated as Pearson's based on nominal returns? Or a Spearman's based on the rank order of the stocks that is determined by the model? Again, we settle the problem after going through some preliminary exploration. We find that both the correlation measures would offer practically identical results for all of our purposes. As a result, we only show Pearson-based ICs in the rest of this paper.

Some key features of our simulation design are summarized in the table below.

**Table 1. Key feature of the simulation design**

| Design factor | Simulation values/assumptions |
| --- | --- |
| Investment Universe size ($N$) | 50, 100, 250, 500, 1000, 3000, 5000 |
| Actual/True IC ($\rho$) range | -0.1, -0.05, -0.04, -0.03, -0.02, -0.01, 0.01, 0.02, 0.03, 0.04, 0.05, 0.06, 0.07, 0.08, 0.09, 0.1, 0.2, 0.3 |
| IC definition | Pearson correlation |
| Distribution of the stock returns | Normal |
| Number of simulation draws for a combination of $N$ and $\rho$ | 1000 |

---

[3] Monthly stock returns in an investment universe are found to be close to a normal distribution; see [8], for example.

With the above setup, we conduct analysis to answer the following questions:

i) When IC is low, how reliably can it be recovered? We answer this "recoverability" question by looking at the relative bias of the estimated ICs.
ii) When IC is low, how low can it go before one can find the underlying model doing more harm than good (or doing worse than a useless/harmless placebo)? We use an alternative model performance measure, Quintile Spread, to help determine acceptable level of ICs. This alternative measure calculates the return difference between the stocks in the first and fifth quintiles of the investment universe.
iii) Can a small negative IC ever be acceptable? This is a derivative question from ii). We can see how a negative IC, even small in magnitude, is related to quintile spread.

When answering these questions, we also try to build the connections between our simulations and some established statistical procedures, verifying the validity of our results.

## 4. Recovery Bias

Assuming that the realized and predicted "normalized" returns follow a bivariate normal distribution whose true correlation is the actual IC (or $\rho$),[4] the estimated IC (or $IC$) is a random variable whose distribution can be characterized by the $\rho$ and the investment universe size ($N$). To assess the behavior of this distribution, we first simulated two random normal variables with correlation $\rho$ and sample size $N$, and then calculated the Pearson correlation between the simulated variables as estimated IC. The relative bias of the estimated IC ($\hat{\rho}$), which can be seen as the opposite to the recoverability of the actual IC, is defined as

$$Relative\ Bias = \frac{\hat{\rho} - \rho}{\rho} \tag{3}$$

The absolute value of the relative bias can be called *IC recovery bias*. Its behavior is plotted in Figure 1 for a subset of $N$ and $\rho$ combinations, and in Figure 2 for a 3-D view of the tri-variate relationship. From these plots, it is clear that the recovery bias decreases both in the universe size and in the magnitude – in absolute term – of the actual IC; the bias would be the largest when the actual IC is very close to zero[5] and the universe size is small.

*Fisher Transformation*

The simulated behavior of the recovery bias observed above can be linked *more formally* to a well-known analysis by Ronald A. Fisher ( [4]). In that analysis, Fisher shows that a properly transformed correlation variable (*IC*) follows approximately a normal distribution [4] as specified below:

$$\frac{1}{2} * ln\left(\frac{1+\hat{\rho}}{1-\hat{\rho}}\right) \sim N(\frac{1}{2} * ln\left(\frac{1+\rho}{1-\rho}\right), \frac{1}{N-3}) \tag{4}$$

The transformation of $\frac{1}{2} * ln\left(\frac{1+\hat{\rho}}{1-\hat{\rho}}\right)$ is called Fisher transformation. When the realized correlation is small with $|IC| < \frac{1}{2}$, as in the cases discussed in this paper, the Fisher transformation is approximately the identify function[6]. As a result, the estimated IC can be approximated as a normal distribution in equation (5):

---

[4] In this paper, "actual" (or "true") IC, is used to refer to the *true predictability* of a model. It is different from two other related terms, "estimated" IC and "realized" IC. An estimated IC is a random variable from a data-generating process that is governed by the actual IC and the universe size in our simulations. A realized IC, as $\alpha_t$ in equation (1) is the correlation between the actual and predicted stock returns realized (observed) at a particular time. We discuss realized IC in section 7.
[5] The bias measure is not defined when the actual IC or $\rho$ is zero.
[6] https://en.wikipedia.org/wiki/Fisher_transformation

$$\hat{\rho} \sim N(\rho, \frac{1}{N-3}) \tag{5}$$

**Figure 1. IC Recovery Bias as a Function of the Actual IC and Universe Size (N): Selected Results**

Panel A - Actual IC=0.01　　　　　　　　　　　Panel B - Actual IC=0.03

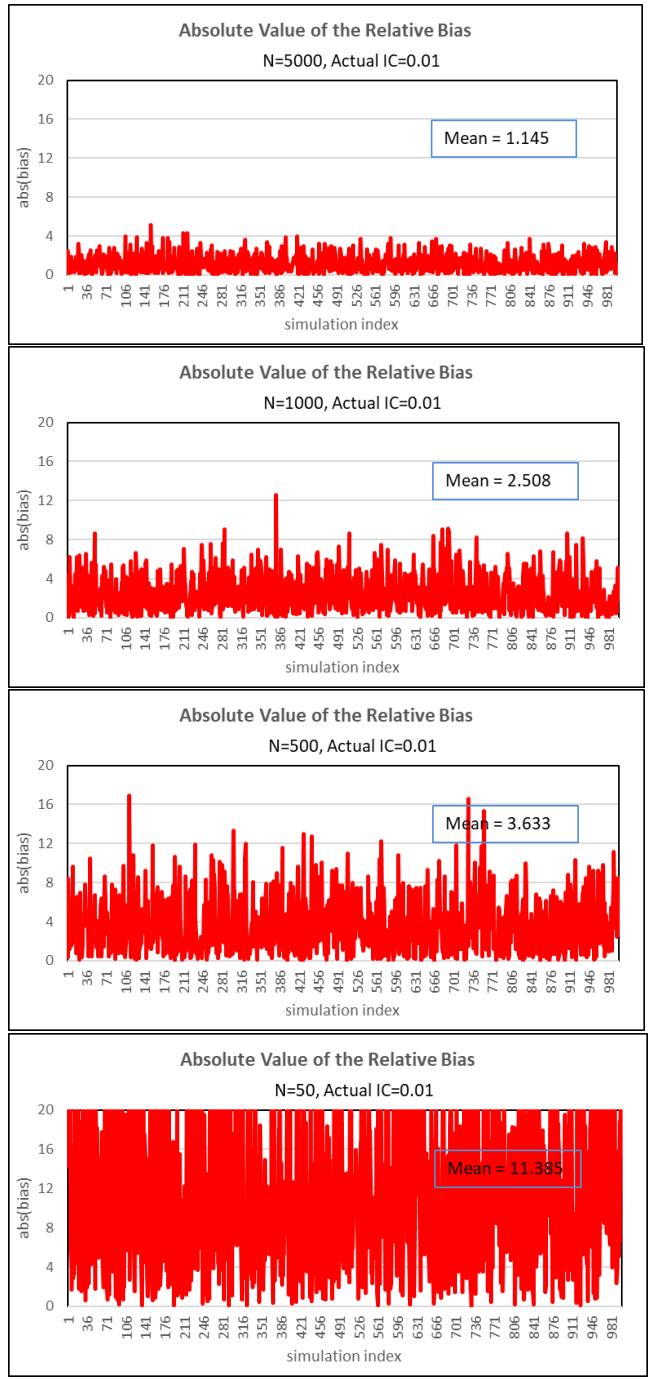
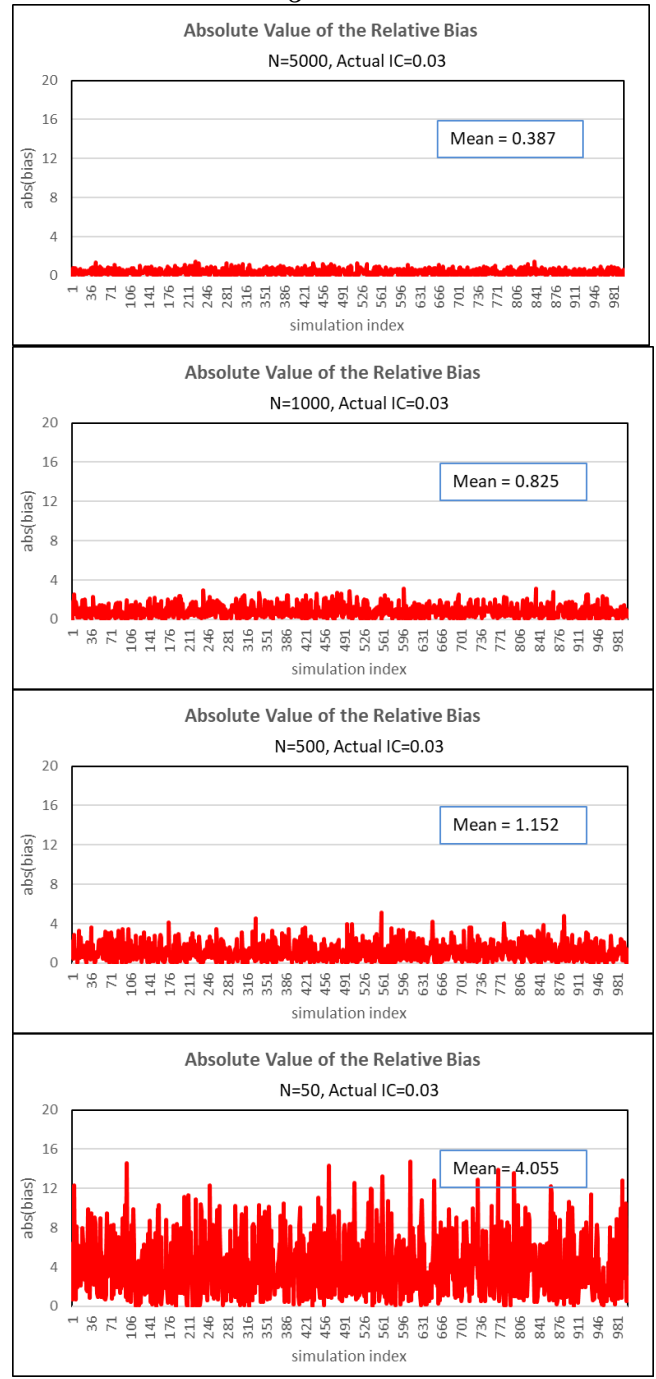

Panel C - Actual IC=0.05         Panel D - Actual IC=0.1

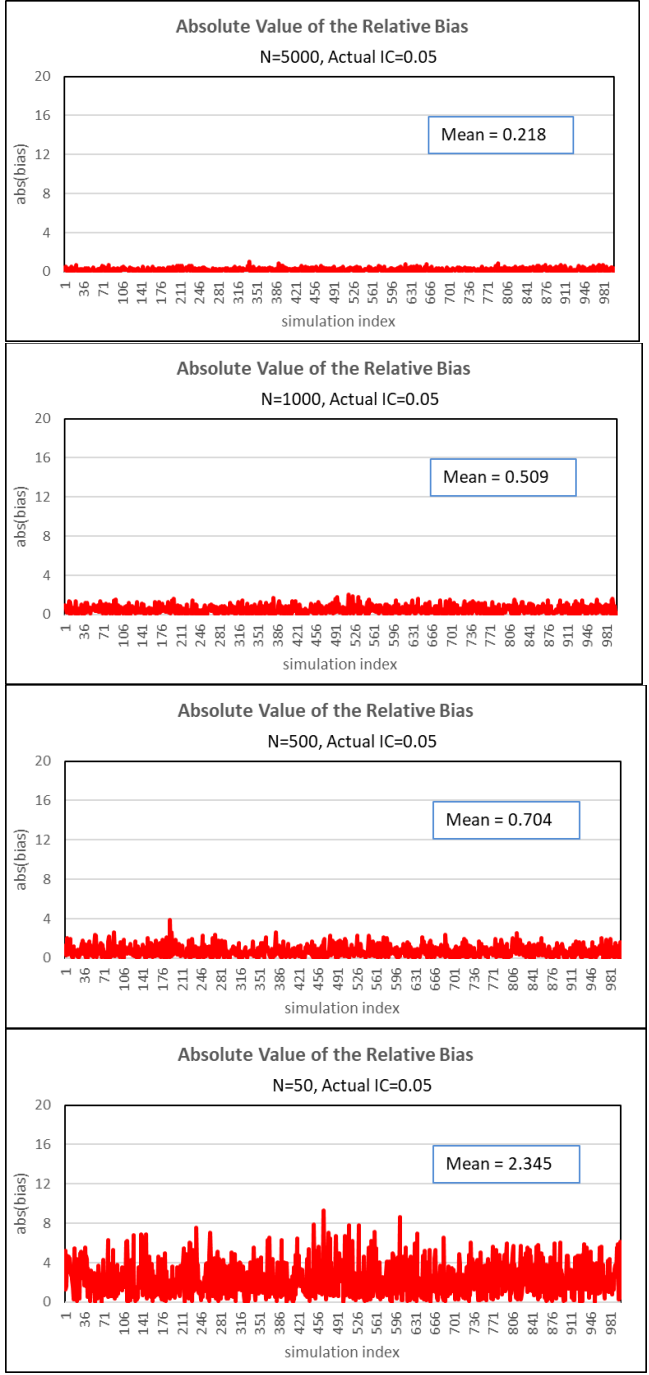
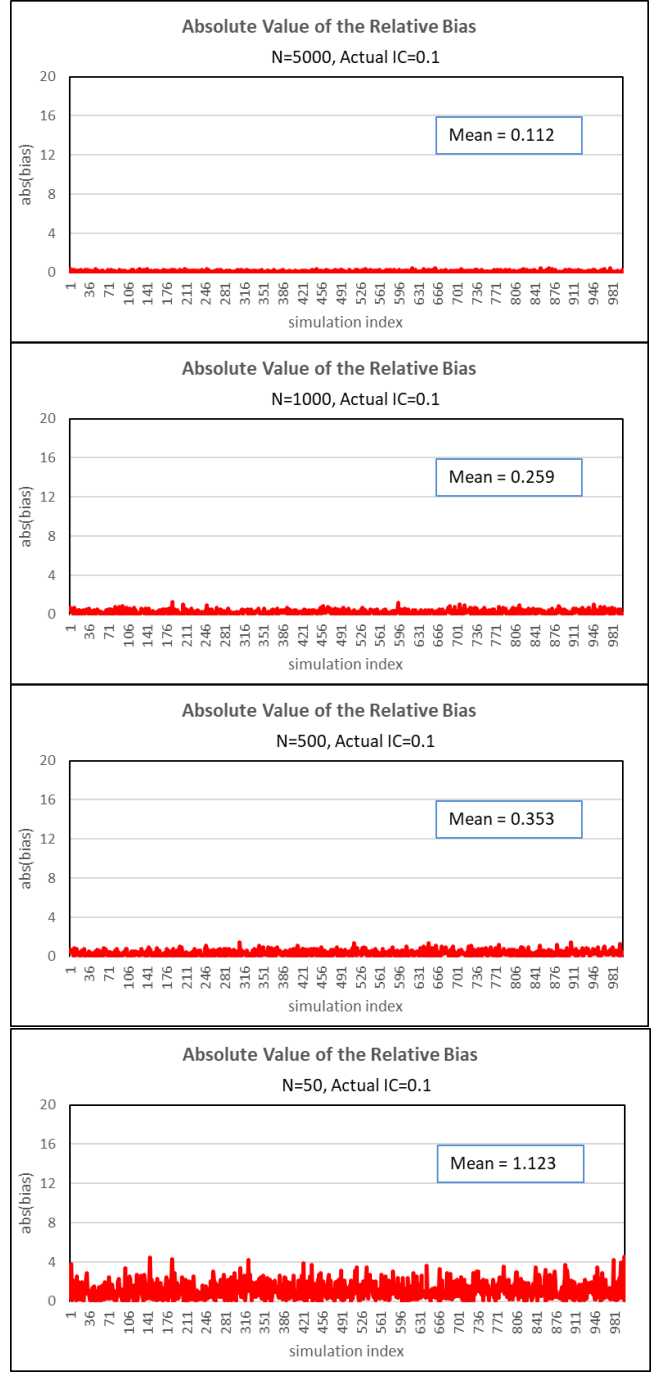

In turn, the relative bias as defined in (3) can be approximated with a normal distribution in equation (6):

$$Relative\ Bias \sim N\left(0, \frac{1}{(N-3)*\rho^2}\right) \qquad (6)$$

Consistent with our results in Figures 1 and 2, the variance of the relative bias indeed decreases with the universe size and actual correlation strength.

The recovery bias (or the absolute value of the relative bias) then approximately follows a half-normal distribution, with an expected value as in equation (7):

$$E(abs(Relative\ Bias)) = E(Recovery\ Bias) \approx \frac{\sqrt{2}}{|\rho|*\sqrt{\pi(N-3)}} \qquad (7)$$

We verified this expected value in our simulation exercise, and found that equation (7) provides a reasonably accurate characterization of the IC behavior studied. The differences between the simulated and the Fisher-expected recovery bias were found small, ranging from -4% to 12% across our simulation exercise. The theoretical IC recovery bias as plotted in Figure 3 matches the pattern as seen in Figure 2.

**Figure 2 IC Recovery Bias as a Function of the Actual IC and Universe Size: A 3-D View[7]**

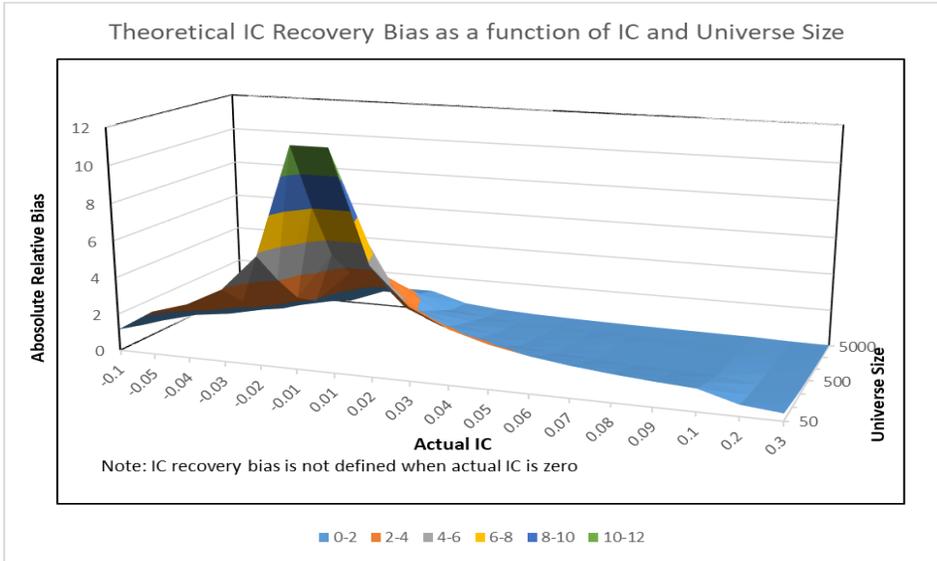

**Figure 3 Theoretical IC Recovery Bias as a Function of the Actual IC and Universe Size: A 3-D View**

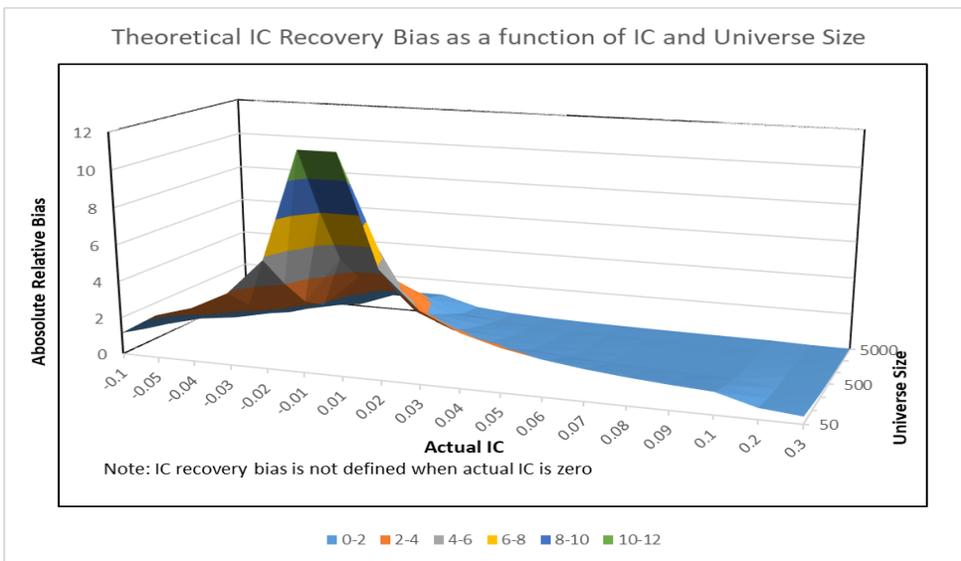

---

[7] In theory, the 3-D view should be symmetric. The asymmetric representation here is completely due to an asymmetric simulation design.

*Confidence Intervals*

The relationship between the estimated IC on the one hand and the actual IC and the universe size on the other can also be seen in the $\rho - N$-specific confidence intervals for the estimation IC. In Table 2, we compare the 95% confidence intervals for a subset of $N$ and $\rho$ combinations, using three procedures: simulation, Equation (4) and Equation (5).

The table suggests that the three procedures can produce practically identical CIs, especially when the universe size is large. In addition, negative estimated ICs are possible even when the actual IC is positive, a result that has long underlined the practical difficulties in using IC as a performance measure of stock selection models.

**Table 2 95% Confidence Intervals for by Actual IC and Universe Size**

| Actual IC ($\rho$) | Universe Size ($N$) | Procedure Used | | |
|---|---|---|---|---|
| | | Simulation | Equation (4) | Equation (5) |
| -0.05 | 50 | (-0.319, 0.249) | (-0.324, 0.232) | (-0.336, 0.236) |
| -0.05 | 500 | (-0.134, 0.034) | (-0.137, 0.038) | (-0.138, 0.038) |
| -0.05 | 5000 | (-0.077, -0.024) | (-0.078, -0.022) | (-0.078, -0.022) |
| -0.01 | 50 | (-0.300, 0.269) | (-0.288, 0.269) | (-0.296, 0.276) |
| -0.01 | 500 | (-0.099, 0.081) | (-0.098, 0.078) | (-0.098, 0.078) |
| -0.01 | 5000 | (-0.039, 0.016) | (-0.038, 0.018) | (-0.038, 0.018) |
| 0 | 50 | (-0.280, 0.260) | (-0.278, 0.278) | (-0.286, 0.286) |
| 0 | 500 | (-0.085, 0.098) | (-0.088, 0.088) | (-0.088, 0.088) |
| 0 | 5000 | (-0.029, 0.029) | (-0.028, 0.028) | (-0.028, 0.028) |
| 0.01 | 50 | (-0.262, 0.295) | (-0.269, 0.288) | (-0.276, 0.296) |
| 0.01 | 500 | (-0.079, 0.100) | (-0.078, 0.098) | (-0.078, 0.098) |
| 0.01 | 5000 | (-0.019, 0.037) | (-0.018, 0.038) | (-0.018, 0.038) |
| 0.05 | 50 | (-0.222, 0.326) | (-0.232, 0.324) | (-0.236, 0.336) |
| 0.05 | 500 | (-0.040, 0.133) | (-0.038, 0.137) | (-0.038, 0.138) |
| 0.05 | 5000 | (0.023, 0.077) | (0.022, 0.078) | (0.022, 0.078) |
| 0.1 | 50 | (-0.168, 0.364) | (-0.183, 0.368) | (-0.186, 0.386) |
| 0.1 | 500 | (0.014, 0.185) | (0.012, 0.186) | (0.012, 0.188) |
| 0.1 | 5000 | (0.073, 0.129) | (0.072, 0.127) | (0.072, 0.128) |

Note: Only selected scenarios are presented here for illustration.

## 5. Quintile Spread, and Acceptable ICs

Quintile spread is an alternative metrics that measures the performance of stock selection models. This metric calculates the return difference between the stocks in the first and fifth quintiles of the investment universe. Based on Equation (1), we can write the theoretical Quintile spread as

$$Quintile\ Spread = E(y|x > 80\ percentile) - E(y|x < 20\ percentile) \qquad (8)$$

Given that the error term in Equation (1) is independent from the predicted return, Equation (8) can be rewritten as

$$Quintile\ Spread = \rho * E(x|x > 80\ percentile) - \rho * E(x|x < 20\ percentile) \qquad (9)$$

Based on the characteristics of the truncated normal distribution[8], the theoretical Quintile Spread can be determined as:

$$Quintile\ Spread = \rho * \left(\frac{\phi(normal.inverse(0.8))}{1-0.8}\right) - \rho * \left(\frac{-\phi(normal.inverse(0.2))}{0.2}\right) \approx 2.80 * \rho \qquad (10)$$

**Table 3 Average Ratio of Quintile Spread to Actual IC from Simulation**

| | Actual IC | | | | | | | | | | | | | | | | | |
|---|---|---|---|---|---|---|---|---|---|---|---|---|---|---|---|---|---|---|
| N | -0.1 | -0.05 | -0.04 | -0.03 | -0.02 | -0.01 | 0.01 | 0.02 | 0.03 | 0.04 | 0.05 | 0.06 | 0.07 | 0.08 | 0.09 | 0.1 | 0.2 | 0.3 |
| 50 | 2.80 | 2.75 | 2.86 | 2.70 | 1.97 | 6.31 | 6.09 | 3.62 | 1.95 | 3.12 | 2.67 | 2.73 | 2.93 | 2.56 | 2.65 | 2.69 | 2.65 | 2.76 |
| 100 | 2.84 | 2.82 | 3.32 | 3.74 | 3.14 | 1.46 | 2.95 | 2.60 | 2.43 | 2.99 | 2.94 | 2.79 | 2.80 | 3.06 | 2.83 | 2.85 | 2.73 | 2.78 |
| 250 | 2.81 | 2.74 | 2.84 | 2.79 | 2.42 | 2.63 | 2.80 | 2.52 | 3.09 | 2.99 | 2.67 | 2.76 | 2.81 | 2.76 | 2.77 | 2.65 | 2.78 | 2.81 |
| 500 | 2.80 | 2.75 | 2.74 | 2.72 | 2.96 | 2.57 | 3.70 | 2.62 | 2.66 | 2.54 | 2.79 | 2.91 | 2.74 | 2.85 | 2.69 | 2.79 | 2.80 | 2.82 |
| 1000 | 2.75 | 2.93 | 2.75 | 2.80 | 2.81 | 2.87 | 2.65 | 2.71 | 2.85 | 2.75 | 2.66 | 2.84 | 2.72 | 2.82 | 2.87 | 2.80 | 2.80 | 2.79 |
| 3000 | 2.80 | 2.84 | 2.79 | 2.72 | 2.76 | 2.99 | 2.59 | 2.61 | 2.69 | 2.78 | 2.77 | 2.77 | 2.78 | 2.79 | 2.84 | 2.78 | 2.80 | 2.80 |
| 5000 | 2.79 | 2.80 | 2.79 | 2.77 | 2.95 | 3.05 | 2.88 | 2.82 | 2.75 | 2.80 | 2.79 | 2.76 | 2.77 | 2.83 | 2.82 | 2.81 | 2.80 | 2.80 |

Note: Values with 10+% errors from the theoretical 2.8 are highlighted

**Table 4 Prevalence of Negative Quintile Spread as a Function of IC and Investment Universe Size**

| | Investment Universe Size (# of Stocks) | | | | | | |
|---|---|---|---|---|---|---|---|
| IC | 50 | 100 | 250 | 500 | 1000 | 3000 | 5000 |
| -0.1 | 0.7390 | 0.8230 | 0.9150 | 0.9810 | 0.9990 | 1.0000 | 1.0000 |
| -0.05 | 0.6310 | 0.6710 | 0.7520 | 0.8480 | 0.9290 | 0.9930 | 0.9990 |
| -0.04 | 0.5740 | 0.6810 | 0.7190 | 0.7610 | 0.8690 | 0.9750 | 0.9960 |
| -0.03 | 0.5770 | 0.6390 | 0.6680 | 0.7150 | 0.7990 | 0.9250 | 0.9680 |
| -0.02 | 0.5310 | 0.5770 | 0.5940 | 0.6620 | 0.7160 | 0.8270 | 0.9000 |
| -0.01 | 0.5420 | 0.5220 | 0.5440 | 0.5790 | 0.6030 | 0.6930 | 0.7460 |
| 0 | 0.5120 | 0.4880 | 0.4880 | 0.4970 | 0.4850 | 0.5100 | 0.5360 |
| 0.01 | 0.4580 | 0.4670 | 0.4580 | 0.4060 | 0.3850 | 0.3220 | 0.2740 |
| 0.02 | 0.4500 | 0.4380 | 0.3920 | 0.3690 | 0.2850 | 0.1800 | 0.1120 |
| 0.03 | 0.4630 | 0.3960 | 0.3060 | 0.2840 | 0.1950 | 0.0700 | 0.0320 |
| 0.04 | 0.3810 | 0.3460 | 0.2690 | 0.2540 | 0.1350 | 0.0270 | 0.0040 |
| 0.05 | 0.3760 | 0.3130 | 0.2380 | 0.1580 | 0.0930 | 0.0060 | 0.0010 |
| 0.06 | 0.3610 | 0.3070 | 0.1940 | 0.1110 | 0.0420 | 0.0060 | 0.0000 |
| 0.07 | 0.3160 | 0.2690 | 0.1560 | 0.0870 | 0.0310 | 0.0000 | 0.0000 |
| 0.08 | 0.3280 | 0.2170 | 0.1270 | 0.0440 | 0.0150 | 0.0010 | 0.0000 |
| 0.09 | 0.2940 | 0.2000 | 0.1040 | 0.0310 | 0.0090 | 0.0000 | 0.0000 |
| 0.1 | 0.2740 | 0.1830 | 0.1010 | 0.0230 | 0.0060 | 0.0000 | 0.0000 |
| 0.2 | 0.1080 | 0.0410 | 0.0030 | 0.0000 | 0.0000 | 0.0000 | 0.0000 |
| 0.3 | 0.0330 | 0.0030 | 0.0000 | 0.0000 | 0.0000 | 0.0000 | 0.0000 |

Note: The numbers in the table are the proportions of scenarios with negative simulated Quintile Spread. Light red = 50%+; Yellow = 1/3 - 50%; Green = below 10%.

---

[8] https://en.wikipedia.org/wiki/Truncated_normal_distribution

This theoretical relationship between Quintile Spread to actual IC was largely confirmed in our simulation exercise, except for scenarios where the universe size is very small (e.g., $N = 50$) and the actual IC is very close to zero; see Table 3.

The theoretical equivalence between Quintile Spread and IC makes Quintile Spread a valuable tool for evaluating the adequacy of IC as a performance measure of stock selection models. To the extent that the value judgment on a model with a low IC can be very elusive, Quintile Spread may help nullify the potential ambiguity about the performance of the model, thereby adding some clarity on acceptable levels of ICs.

For example, if a stock selection model is used by an investment manager properly (e.g., she longs the stocks in the highest quintile and/or shorts the stocks in lowest quintile), in order for the model to add any value, the Quintile Spread of the model must be at least positive. This intuitive decision rule should help determine the thresholds for acceptable ICs.

In Table 4, we present the prevalence of negative quintile spread from our simulation exercise by actual IC and universe size. As expected, the stronger the IC signal, the less likely the negative quintile spread (i.e., the numbers in the table decreases as IC increases); the larger the investment universe, the more consistent this Quintile Spread-IC relationship (e.g., the prevalence numbers in the last two columns are closer to either 100% or 0%). In general, the table suggests that an acceptable IC for a value-adding stock selection model may be determined by the size of the investment universe and the size of the expected (positive) quintile spread.

Table 4 also suggests that any model with a negative IC would not be a *good* model – Regardless of the universe size, the quintile spread would be negative for such a model. While this result is not surprising, it clearly sets a theoretical lower bound for the ICs of all permissible models.

## 6. Dynamics of Realized ICs

The discussion so far is focused on the behavior of IC at a given point of time – how a true IC can be recovered (revealed) in estimated ICs and how an IC manifests model performance. A stock selection model, however, is often used to support an investment strategy across time. Each time the model is used, a realized IC ($\rho^*$) is obtained, measured against the random actual returns in the investment universe. Consequently, the realized IC itself is a random variable. In fact, the realized ICs has been shown to be volatile ( [5]), and follow a normal distribution over time ( [6]). For simplification, we assume that the realized ICs are independent, with no serial correlation.[9]

To help understand the dynamics of the realized ICs, consider the following simple mechanism:

$$\rho^*_{t,N} = \rho_0 + e_N + e_t \qquad (11)$$

Here, the realized IC obtained at time *t* from a model that covers an investment universe $N$, $\rho^*_{t,N}$, is assumed to be determined by three components: a constant underlying true IC of the model ($\rho_0$), an error term $e_N$ that can be explained by universe size in Equation 5, and a time-specific residual error $e_t$, which is independent of $e_N$.

In this setup, the realized ICs may be seen as an orthogonal combination of a static procedure ($\rho_0 + e_N$), which we have studied earlier, and a time-varying process, $e_t$.

---

[9] In reality, the realized ICs may have serial correlation since the predictors (or factors) used in a model are often impacted by market cycle. This may affect the estimate of the standard deviation of the mean IC and – consequently – hypothesis testing related to the ICs ( [9]). However, the serial correlation of realized ICs is found to be small ( [6]). If the serial correlation were strong, portfolio managers could buy the recent top-performing factors and sell poor-performing factors to gain profit, making the serial correlation to vanish.

The standard deviation of the realized ICs can then be calculated as:

$$std\left(\rho^*_{t,N}\right) = \sqrt{std(e_t)^2 + std(e_N)^2} = \sqrt{std(e_t)^2 + 1/(N-3)} \tag{12}$$

In Table 5, we report the standard deviations of the realized ICs from some industry studies ( [5] and [7]), and decompose the total standard deviations into the static and time-varying components – using equation (12).

**Table 5 Realized ICs and the Standard Deviation Decomposition for Selected Industry Studies**

| Universe Size (N) | Model/Factor | Study period | Average IC ($\rho^*_{t,N}$) | $std\left(\rho^*_{t,N}\right)$ | $std(e_N)$ | $std(e_t)$ | $std(e_t)$ % | Minimal T |
|---|---|---|---|---|---|---|---|---|
| 1000 | Book to price | 1978:12-2009:8 | 0.011 | 0.125 | 0.032 | 0.121 | 96.7 | 349 |
| 2000 | Book to price | 1978:12-2009:8 | 0.013 | 0.095 | 0.022 | 0.092 | 97.2 | 144 |
| 3000 | Book to price | 1978:12-2009:8 | 0.012 | 0.096 | 0.018 | 0.094 | 98.2 | 173 |
| 1000 | Cash flow to price | 1978:12-2009:8 | 0.023 | 0.085 | 0.032 | 0.079 | 92.8 | 37 |
| 2000 | Cash flow to price | 1978:12-2009:8 | 0.033 | 0.094 | 0.022 | 0.091 | 97.1 | 22 |
| 3000 | Cash flow to price | 1978:12-2009:8 | 0.031 | 0.089 | 0.018 | 0.087 | 97.9 | 22 |
| 1000 | Earnings to Price | 1978:12-2009:8 | 0.009 | 0.132 | 0.032 | 0.128 | 97.1 | 582 |
| 2000 | Earnings to Price | 1978:12-2009:8 | 0.025 | 0.12 | 0.022 | 0.118 | 98.2 | 62 |
| 3000 | Earnings to Price | 1978:12-2009:8 | 0.023 | 0.122 | 0.018 | 0.121 | 98.9 | 76 |
| 1000 | Sales to Price | 1978:12-2009:8 | 0.017 | 0.109 | 0.032 | 0.104 | 95.7 | 111 |
| 2000 | Sales to Price | 1978:12-2009:8 | 0.017 | 0.086 | 0.022 | 0.083 | 96.6 | 69 |
| 3000 | Sales to Price | 1978:12-2009:8 | 0.016 | 0.087 | 0.018 | 0.085 | 97.8 | 80 |
| 1000 | 12-Month Momentum | 1978:12-2009:8 | 0.031 | 0.19 | 0.032 | 0.187 | 98.6 | 102 |
| 2000 | 12-Month Momentum | 1978:12-2009:8 | 0.039 | 0.135 | 0.022 | 0.133 | 98.6 | 32 |
| 3000 | 12-Month Momentum | 1978:12-2009:8 | 0.037 | 0.144 | 0.018 | 0.143 | 99.2 | 41 |
| 1000 | Share Repurchase | 1978:12-2009:8 | 0.013 | 0.075 | 0.032 | 0.068 | 90.6 | 90 |
| 2000 | Share Repurchase | 1978:12-2009:8 | 0.021 | 0.068 | 0.022 | 0.064 | 94.4 | 28 |
| 3000 | Share Repurchase | 1978:12-2009:8 | 0.02 | 0.067 | 0.018 | 0.064 | 96.2 | 30 |
| 1000 | Percent Short | 1978:12-2009:8 | 0.017 | 0.134 | 0.032 | 0.130 | 97.2 | 168 |
| 2000 | Percent Short | 1978:12-2009:8 | 0.033 | 0.095 | 0.022 | 0.092 | 97.2 | 22 |
| 3000 | Percent Short | 1978:12-2009:8 | 0.03 | 0.1 | 0.018 | 0.098 | 98.3 | 30 |
| 7500 | Multi-factor | 1997:1–2011:12 | 0.045 | 0.086 | 0.012 | 0.085 | 99.1 | 10 |

Note: Minimal T means the minimum number of time periods needed for average IC to be significantly higher than 0 at 5% level. Data sources: [5] and [7].

It is clear from the table that the realized ICs are volatile. Of about the two-dozen models covered, with an exception for the Multi-factor model in the last row, the standard deviations of the realized ICs are at least three times as large as the means (the max is 11 times), suggesting that a symmetric IC distribution would put a significant portion of the ICs in the negative territory. More importantly, the time-varying component dominates the static component in driving the IC changes across time, accounting for about 95% of the total variability. This result clearly underlines the great difficulties in relying on the realized ICs to monitor ongoing performance of stock selection models, a problem that we attempt to address next.

## 7. Ongoing Performance Monitoring Using Realized ICs

Average IC test

We propose two alternative procedures for using realized ICs to monitor the ongoing performance of a stock selection model. The first focuses on the behavior of the average IC. As shown in Section 7, the standard deviations of the realized ICs are very large relative to the mean; it would be extremely difficult to draw any conclusion about the underlying IC if one only looks at the current or otherwise a limited number of realized ICs. The performance of the underlying model should thus be evaluated based on the realized ICs over a reasonable period of time (i.e., data window).

Depending on data availability, and following some convenient conventions, one may define a *reasonable* period of time as last 6 months, last 12 months, last 24 months or last 36 months.[10] An average realized IC can be calculated for any of these data windows. A long data window tends to lead to a narrower confidence interval for the realized IC, whereas a short window may be more appropriate for detecting model performance deterioration due to market volatility. In general, careful judgment should be used when choosing an appropriate data window for the average IC analysis.

The average IC so obtained can be used to form a hypothesis test, with the null hypothesis being that the underlying IC is $\rho_0$. The standard deviation of the IC for this test can be determined either from all the historical data, or from the specific data window chosen.

The choice on $\rho_0$ can be tricky, although some practical guidance may be available. For example, one may choose the estimated IC from the initial model development as the underlying IC. In this case, the hypothesis testing aims to monitor whether the model remains as relevant as the originally developed. The underlying IC can also be an expected IC set by a manager, following a process similar to what we have discussed in section 5. Overall, in order to maintain an effective procedure for ongoing monitoring of model performance, the model performance goal must be clearly set.

**Table 6 95%-Critical Value for the Average Realized IC in One-sided Hypothesis Test: Selected Data Windows**

| Underlying IC ($\rho_0$) | std$\left(\rho^*_{t,N}\right)$ | Data Window: 12 Months | Data Window: 24 Months | Data Window: 36 Months |
|---|---|---|---|---|
| 0 | 0.06 | -0.028 | -0.020 | -0.016 |
| 0 | 0.08 | -0.038 | -0.027 | -0.022 |
| 0 | 0.1 | -0.047 | -0.034 | -0.027 |
| 0.01 | 0.06 | -0.018 | -0.010 | -0.006 |
| 0.01 | 0.08 | -0.028 | -0.017 | -0.012 |
| 0.01 | 0.1 | -0.037 | -0.024 | -0.017 |
| 0.05 | 0.06 | 0.022 | 0.030 | 0.034 |
| 0.05 | 0.08 | 0.012 | 0.023 | 0.028 |
| 0.05 | 0.1 | 0.003 | 0.016 | 0.023 |

Note: The standard deviations are scaled across data windows by using the square-root of the time rule. We focus on one-sided test here because the main concern of this test is about model performance deterioration.

---

[10] Without any loss of generalizability, we assume here that the model performance is evaluated and monitored on a monthly basis.

In Table 6, we show the 95%-critical values of the average realized IC for three alternative data windows under different assumptions on the underlying IC ($\rho_0$) and the standard deviation of the ICs. For example, in order to maintain an underlying IC of 0.01, the average IC obtained from a 12-month data window should not be lower than -0.028 if the standard deviation of the IC is 0.08. It should be noted that the realized ICs are assumed to be independent. If serial correlation was considered, the critical values could be even lower.

*Binomial test*

The second procedure that one may use to monitor model performance on an ongoing basis builds upon the average IC test described above. It takes the outcomes (i.e., rejecting or accepting the null hypothesis) from the monthly hypothesis testing exercise, and evaluates the outcomes across time.

To illustrate, suppose the monthly hypothesis testing is based on a significance level of 5%. If the null hypothesis is true, there is a 5% chance of rejecting the hypothesis each month. Assume that all the tests are i.i.d., then the number of rejects in a year would follow a binomial distribution. The probability of having 3 or more rejects over the 12 month window is about 2%, meaning that the maximum acceptable number of rejects over the window is 2 at a 2% level of error tolerance. Some additional results from this binomial process are in Table 7.

**Table 7 Binomial Distribution for the Number of 5%-Rejects in 12 Month**

| Number of Rejects | Probability (or Error tolerance level) |
|---|---|
| 1 or more | 46% |
| 2 or more | 12% |
| 3 or more | 2% |
| 4 or more | 0.2% |

*Power of the hypothesis testing*

When conducting a hypothesis test, the probability of rejecting a true null hypothesis (or the significance level) is not the only concern. The power of the test, or the probability of rejecting a false null hypothesis, should also be considered. The power of the one-sided average IC test can be written as

$$power = \Phi(\frac{IC_0 - IC_1}{\frac{\sigma_{IC}}{\sqrt{T}}} - Z_\alpha) \tag{13}$$

For the binomial test, the power can be written as

$$power = 1 - Binomial(S, T, \Phi\left(\frac{IC_0 - IC_1}{\sigma_{IC}} - Z_{\alpha m}\right)) \tag{14}$$

Here, $T$ is the number of periods in the data window; $\Phi$ is the standard normal cumulative distribution function; $IC_0$ is the underlying IC level under null hypothesis, and $IC_1$ is the alternative underlying IC; $\sigma_{IC}$ is the standard deviation of realized IC; $\alpha m$ is the significance level for the monthly hypothesis testing; S is the maximum number of rejects acceptable for the binomial test.

The significant level and the power of a hypothesis test are related. While this relationship is self-evident as in equation (13), the significance level for the binomial test, $\alpha$, can be determined as below:[11]

$$\alpha = 1 - Binomial(S, T, \alpha m) \tag{15}$$

---
[11] Note that this power estimate $\alpha$ should not be confused with the $\alpha_t$ used in equation (1) earlier.

The formulas indicates that the power of the binomial test would increase with the number of observations in the data window, increase with the difference in underlying IC (i.e., $IC_0 - IC_1$), and decrease with the standard deviation of IC.

In Table 8, we show the power of the two hypothesis testing procedures proposed above, for two significance levels (5% and 2%) and based on various assumptions on the difference in underlying IC and on the IC standard deviation. For example, for a model with an IC reduction of 0.05 (e.g., $IC_0 - IC_1 = -0.05$) and an IC standard deviation of 0.08, the average IC test using a 0.012 critical value would have a 69.9% probability of detecting the model deterioration based on a 12-month data window. However, for a minor IC reduction such as $IC_0 - IC_1 = 0.01$, the power of the test can be low for both procedures, suggesting a statistical difficulty in detecting minor model performance deterioration over a short data window.

**Table 8 Power of the 12 month hypothesis tests**

| | | Average IC test | | Binomial test (keep $S = 2$, adjust $\alpha m$) | |
|---|---|---|---|---|---|
| $IC_0 - IC_1$ | $\alpha$ / $\sigma_{IC}$ | 5% | 2% | 5% | 2% |
| 0.01 | 0.06 | 0.143 | 0.070 | 0.105 | 0.047 |
| 0.01 | 0.08 | 0.113 | 0.053 | 0.088 | 0.038 |
| 0.01 | 0.1 | 0.097 | 0.044 | 0.079 | 0.034 |
| 0.03 | 0.06 | 0.535 | 0.374 | 0.327 | 0.188 |
| 0.03 | 0.08 | 0.365 | 0.225 | 0.225 | 0.118 |
| 0.03 | 0.1 | 0.272 | 0.155 | 0.175 | 0.087 |
| 0.05 | 0.06 | 0.893 | 0.798 | 0.654 | 0.476 |
| 0.05 | 0.08 | 0.699 | 0.544 | 0.446 | 0.281 |
| 0.05 | 0.1 | 0.535 | 0.374 | 0.327 | 0.188 |

## 8. Conclusions

Realized information coefficients are small in magnitude and volatile across time, making it difficult to use the metric as a performance measure of stock selection models. In this paper, we have attempted to understand the behavior of the realized ICs, both statically and dynamically.

In statics, where the ICs are evaluated at a given point of time, our simulation study showed that the recovery bias decreases in the investment universe size and decreases in the magnitude of the actual IC, attesting to the difficulties in evaluating small ICs in the investment field. This result was found to be consistent with a statistical procedure developed by Fisher ( [4]). Our simulation study also confirmed that simulated ICs can be directly linked to quintile spread; this connection can help investment managers determine acceptable ICs for their models.

In dynamics, we developed a simple model to understand the realized ICs across time, decomposing the ICs into a static component and a time-varying counterpart. Based on data from industry studies, we found that the realized ICs are indeed very volatile, with the volatility predominantly driven by the time-varying component. This suggests that any snapshot/static views of the realized ICs would not be useful for evaluating model performance.

Finally, we proposed two ongoing performance monitoring procedures that use the realized ICs, and discussed the power of these testing procedures. Through illustration, we showed how these procedures, if carried out carefully, may help detect the potential deterioration of model performance over time.